\documentclass[10pt,a4paper]{article}
\usepackage{graphpap,epsfig,amssymb,graphicx,textcomp,amsmath,xcolor}

\begin{document}

\begin{center}

{\Large {\bf Compensation in trilayered anisotropic 6-state clock model}}\\

\vskip 0.3cm

{\it Olivia Mallick$^{a,*}$ and Muktish Acharyya$^{b,+}$}\\

\vskip 0.3cm

{\it $^a$Department of Physics of Complex Systems, S. N. Bose National Centre for Basic Sciences, Kolkata-700106,India}\\

{\it $^b$Department of Physics, Presidency University, 86/1 College Street, Kolkata- 700 073, India}

\end{center}

\begin{abstract}
 The equilibrium behaviours of a trilayered 6-state clock model have been investigated by Monte Carlo simulation. The intralayer interaction is considered ferromagnetic, whereas the interlayer interaction is antiferromagnetic. Periodic  boundary conditions(PBC) are applied in XY plane and open boundary condition (OBC) is applied along Z-direction. The thermodynamic behaviours of sublattice magnetisation, total magnetisation, magnetic susceptibility and the specific heat are studied as functions of the temperature. The interesting compensation phenomenon, when the total magnetisation vanishes keeping the nonzero sublattice magnetisations, has been observed. The compensation temperature and the critical temperature are found to depend on the strength of single-site anisotropy. The comprehensive phase diagram is shown in the temperature-anisotropy plane. 
The larger systems show the growth of the height of the peak of the susceptibility near the critical temperature.
\end{abstract}

\vskip 4cm


\vskip 0.2cm


\vskip 3cm

\noindent {$^*$E-mail:mallickolivia0@gmail.com}\\
\noindent {$^+$E-mail:muktish.physics@presiuniv.ac.in}

\newpage

%\begin{twocolumn}

%\tableofcontents
\section{Introduction}

The ferromagnetic models showing the unusual ordered phase even in absence of conventional long-range ferromagnetic
order is a remarkable discovery\cite{B1,B2}. The continuous symmetric (SO(2)) planar ferromagnetic model shows this
behaviour in two dimensions with vortex-antivortex special kind of ordered state\cite{KT1,K}. The unconventional
metastable behaviour and phase transition in this kind of model has drawn the attention of the researchers in last 
four decades. The celebrated Berezinskii-Kosterlitz-Thouless (BKT)\cite{BKTart} transition is now under the spotlight of modern research in condensed matter physics
and statistical mechanics. \textcolor{blue}{It is a transition from bound vortex-antivortex pairs at low temperatures to unpaired vortices and anti-vortices at some critical temperature. The BKT phase gives rise to a special kind of phase where the conventional long range
ferromagnetic order is absent. This is observed in two dimensions of SO(2) symmetric models. The planar ferromagnetic XY
model in two dimensions shows BKT transitions. The two dimensional superconductor\cite{BKTsc}, liquid crystalline systems
\cite{liqc} also shows BKT
transitions.}

{\it Is this kind of unusual phase the property of continuous symmetry only?} This question has triggered to investigate 
the discrete symetric q-state clock model, where the continuous symetric XY model can be restored in the limit
$q \to \infty$. The properties of q-state clock model has been investigated\cite{clock1} for q=4, 5 and 6 in two dimensions by Monte Carlo renormalization
group analysis to study the phase transitions. The Kosterlitz-Thouless type intermediate phase has been found\cite{crit-clock6} in two dimensional
6-state clock model by Monte Carlo simulations. The KT type phase transition has been found to occur in the range between ${{k_BT} \over {J}}=0.68\pm0.02$ and 
${{k_BT} \over {J}}=0.92\pm0.01$. However, it is confirmed later that there exists a Kosterlitz-Thouless (KT) like phase in the ferromagnetic 6-state clock model on the square lattice through studies of the interfacial free energy estimated by Monte Carlo simulations and the authors estimated\cite{phase-clock-2d} that lower and upper transition temperatures are 0.75 and 0.90 respectively. Interestingly, in a recent study\cite{ml-clockq}, the machine-learning detection of BKT transition was found
in q-state clock model with the neural network based estimates of the two critical points as ${{k_BT} \over {J}}=0.410$ and 
${{k_BT} \over {J}}=0.921$.

The classical 2D clock model is known to have a critical phase with Berezinskii–Kosterlitz–Thouless (BKT) transitions. These transitions have logarithmic corrections which make numerical analysis difficult. Hence, the classical 2D 6-state clock model has been studied \cite{bkt-clock6} by mapping it to the quantum self-dual one dimensional 6-state clock model and confirmed that the self-dual point has a precise numerical agreement with the analytical result. The degenreracy of the excited states at the self
dual point has also been arugued from the perspective of the effective field theory.

The thermodynamic behaviours of the clock model has been studied in different kinds of lattices to investigate the effects of
geometrical connections and interactions. The BKT transition has been observed\cite{phase-clock6} in 6-state clock model
on the rewired squared lattice studied by Monte Carlo simulation with Wang-Landau algorithm and estimated the lower and upper critical temperatures. For pure case, i.e., the lattice with average number of bonds equal to 2, ${{k_B T1} \over {J}}$ and
${{k_B T2} \over {J}}$, are roughly estimated around 0.55 and 1.15 respectively (these results may be compared to those reported in the
references \cite{crit-clock6} \cite{phase-clock-2d} and \cite{ml-clockq}). The BKT transition has been studied\cite{bkt-clock62d} recently, in 6-state 
clock model on different kinds (triangular, kagome, honeycomb and diced) of two-dimensional lattices.
The two-dimensional q-state clock model on random lattices has been investigated {clock-random} via numerical simulation. Interestingly, for q=4 a single phase transition has been reported. Whereas, for q=5 and q=6 multiple (double) phase transitions
are observed from the two peaks of the specific heat.
 The lower and upper critical temperatures are found to depend significantly on the structure of the lattice. The anisotropic (with anisotropy parameter $\alpha$) 6-state clock model has been investigated\cite{anis-clock6} by transfer matrix renormalization group method to explore
the phase transition in two dimensional square lattice. Three distinct phase are identified and a comprehensive phase diagram
is drawn. They observed a single discontinuous phase transition and two BKT phase transitions meeting in the tricritical point.
The tricritical point is estimated at critical anisotropy $\alpha_c = 0.21405$. 
The antiferromagnetic 5-state clock-model has been studied on the triangular lattice recently\cite{clock-antiferro} with second nearest neighbour interaction and the phase diagram is drawn.

The thermodynamic behaviours of clock model in layered structured lattices are not yet reported in the well developed literature
of this specific model. However, the behaviours of layered Ising and Blume-Capel models are well studied. Let us briefly mention
a few selected studies on layered Ising and Blume-Capel models. The main interesting phenomenon in the thermodynamic behaviours
of layered Ising and Blume-Capel models is the {\it compensation} transition. The compensation temperatures and magnetic susceptibility of a mixed ferro-ferrimagnetic ternary alloy have been studied\cite{dely} by Monte Carlo simulations and observed
outstanding behaviour of magnetic susceptibility near the compensation and critical temperature. The compensation temperature
has been found\cite{bobak} to depend of the anisotropy in a mixed spin (spin-${{1} \over {2}}$ and spin-$1$) Heisenberg ferrimagnet 
on a simple cubic lattice, studied by Oguchi approximation. The anisotopy dependent spin compensation temperature has also been
noticed\cite{hadey} in mixed spin (spin-${{3} \over {2}}$ and spin-${{5} \over {2}}$) ferrimagnetic Ising model studied by
mean field approximation. In this case, the possibility of multiple compensation temperature was indicated. The impurity- concentration dependent compensation\cite{comp-ma1} and the anisotropy dependent compensation\cite{comp-ma2} was also noticed in the trilayered Ising and Blume-Capel models, respectively. The meanfield and effective field theory has been employed \cite{diaz-comp} to study the interaction dependent compensation in spin-${{1} \over {2}}$ Ising trilayered ferrimagnetic system. The 
phase diagram (compensation and no-compensation) was drawn in parameter space. The antiferromagnetic systems showed the compensation behaviours in Ising mixed spin (spin-${{3} \over {2}}$ and spin-${{5} \over {2}}$) core/shell nanowire\cite{comp-anti1} and in mixed spin (spin-${{7} \over {2}}$ and spin-${1}$) antiferromagnetic ovalene nanowire\cite{comp-anti2}. The
compensation was found\cite{comp-bilayer}
Ising mixed spin (spin-${{3} \over {2}}$ and spin-${{1} \over {2}}$) graphene bilayered ferrimagnet. The effects of 
exchange interactions, crystal field and external magnetic field on the compensation temperature were studied \cite{comp-bilayer} by Monte Carlo 
simulation. \textcolor{blue}{The parameter dependent compensation temperature has been noticed\cite{extra1} in the Monte Carlo study 
of mixed spin (spin-${{3} \over {2}}$ and spin-${{5} \over {2}}$) Ising model. However, no such compensation was noticed\cite{extra2}
in the study of such kind of mixed spin system in the presence of magnetic field and with ferri and ferromagnetic exchange interactions. The multilayered (composed of two blocks: the upper block A with spin-${{5} \over {2}}$ and the bottom one B with spin-${{3} \over {2}}$)
system has been studied\cite{extra3} by Monte Carlo simulation to investigate the critical and compensation behaviour. The compensation behaviour
has been investigated\cite{extra4} in mixed spin (spin-${{3} \over {2}}$ and spin-${{1} \over {2}}$) Ising ferrimagnetic system with site dilution by nonmagnetic impurities. The compensation temperature has been found to depend strongly on the dilution and
interestingly a critical dilution was recorded above which the compensation disappeared.}

The evidence of engineering the compensation in ferrimagnetic iron garnet by tuning the ratio of in-plane and out of plane magnetisations can be found in \cite{comp-field}. An interesting spin compensation and first-order reentrant behaviour
has been observed\cite{re-comp} in mixed spin (spin-1 and spin-2) Ising ferrimagnetic system studied by Monte Carlo simulation.

{Can the discrete symmetric planar models (6-state clock model) show the compensation behaviour?} Being motivated by this question, we have thoroughly investigated the trilayered 6-state clock model by Monte Carlo simulation. The results are reported
with diagrams in this manuscript. This manuscript is organised as follows: The model and the simulation scheme are discussed
in Section-2. The numerical results are reported with diagrams in Section-3 and the paper ends with concluding remarks in
Section-4.

\label{intro}

%%%%%%%%%%%%%%%%%%%%%%%%%%%%%%%%%%%%%%%%

%\paragraph*{Introduction.-}

\vspace{0.2cm}
\section{Model and numerical details }
\textcolor{blue}{We consider a classical trilayered (A-B-A type) 6-state clock model defined on a simple cubic lattice, where each of the three layers consists exclusively of either type-A or type-B atoms. These different types of atoms (in each layer)
are distinguished by different intralayer ferromagnetic interaction strengths. The magnetic interactions and anisotropic contributions to the energy of the system are captured by the following Hamiltonian: }

\begin{equation}
 \textcolor{blue}{\mathcal{H}= -J_{ferro}^{AA} \sum_{\langle ij \rangle} \vec{S_i} \cdot \vec{S_j} -J_{ferro}^{BB} \sum_{\langle ij \rangle} \vec{S_i} \cdot \vec{S_j}-J_{antiferro}^{AB} \sum_{\langle ij \rangle} \vec{S_i} \cdot \vec{S_j} -D \sum_{i}[({S_i}^x)^2-({S_i}^y)^2].}
\end{equation}

\textcolor{blue}{Here, $\vec{S_i}=(S_i^x, S_i^y)$ represents the classical two-dimensional spin vector at lattice site $i$, which can point in one of six discrete directions in the $xy$-plane, consistent with the 6-state clock model. These discrete orientations are defined by:}
 \begin{equation*}
     S_i^x=\cos{\theta_i}, S_i^y=\sin{\theta_i},
 \end{equation*}

 \noindent where $\theta_i$ can assume any value discretely,
     \begin{equation*}
         \theta_i=\frac{2\pi}{6}n_i, [n_i=0,1,2..5]
     \end{equation*}
 \textcolor{blue}{The first term in Equation-1 accounts for the intralayer ferromagnetic interactions between nearest neighbour spins $\langle ij \rangle$ within the same atomic plane (for top and bottom layers of A-type atom), with a positive coupling exchange $J_{ferro}^{AA} (>0)$. The ferromagnetic contribution arising from middle layer (B-type of atom) is denoted by the second term, where
 the nearest neighbour ferromagnetic interaction strength is denoted as $J_{ferro}^{BB} (> 0)$.
  The third term represents interlayer antiferromagnetic coupling between the adjacent layers, involving nearest neighbour sites $\langle ij \rangle$ across the layers, governed by a coupling constant $J_{antiferro}^{AB} (<0)$. The fourth term introduces an anisotropy $D$ (measured in the unit of absolute value of $J_{ferro}^{AB}$) , which destroys the discrete rotational symmetry in the $xy$-plane and favours the spin-alignment in specific direction. Positive values of $D$ prefer specific spin orientations, thereby reinforcing the discrete six-fold symmetry and enhancing the stability of the selected clock states. We impose periodic boundary conditions (PBC) in the x and y directions within each layer and along the stacking direction z, we adopt free boundary conditions to capture the finite thickness of the trilayered structure.}

 \textcolor{blue}{To study the compensation behaviour in the trilayered 6-state clock model, we employ the Monte Carlo simulation technique based on the Metropolis algorithm with random single-spin updates. The system consists of three stacked square lattices, each consisting of $L\times L$ sites, resulting in a total of $3L^2$ spins. For the majority of our simulations, we consider a system size of $L=128$, corresponding to $128\times 128$ spins per layer. In order to examine finite-size effects, additional simulations are performed for various lattice sizes (32, 64, 128 and 256).} 
 
\textcolor{blue}{ In our simulation, initially, spin configurations are generated at high temperature with randomly assigned orientations chosen from the six allowed spin states in the clock model. This corresponds to the high temperature paramagnetic phase. A high value of the temperature ($T$) is assigned and the system is allowed to evolve using Metropolis single spin-flip scheme. We have employed 
 the random updating rule. A site is chosen randomly from $3L^2$ sites. The spin of this randomly chosen site has to be updated.
 We have calculated the energy of the system with the present spin orientation of the randomly chosen site. Now a trial direction of that particular spin is chosen (randomly from all possible 6 orientations). The change in energy ($\delta E$) has been calculted. Whether the trial direction will be selected or not, that has been decided by  Metropolis acceptance probability\cite{binder}: $P_f= Min [\exp{(-\frac{\delta E}{k_BT}}),1]$. \textcolor{red}{$\delta E$}  is the change in energy
 $E$ (measured in the unit of $J_{ferro}^{AB}$) associated with the proposed spin flip, and $k_B$ is the Boltzmann constant. The temperatures $T$ is measured in the unit of 
${{J_{ferro}^{AB}} \over {k_B}}$. In this way, $3L^2$ such random updating are done, which constitutes one Monte Carlo Step
per site (MCSS). At any fixed temperature, the system is allowed to run for 1.5$\times 10^5$ MCSS, which is sufficient to 
achieve the steady value (equilibriation). The thermodynamic quantities are calculated by averaging over further 1.5$\times 10^5$ MCSS. As a whole, the total length of the simulation is 3.0$\times10^5$ MCSS.}
 
 To characterize the thermodynamic behaviour of the 6-state clock model, we calculate several key observables that capture the collective magnetic ordering of the system. The sublattice magnetisation provides insight into the local symmetry breaking and is defined for each sublattice $k\in{1,2,3}$ as the average spin components in the x and y directions:
 \begin{equation*}
     m_k^x=\frac{1}{L^2}\sum_{i,j}S^x(i,j,k) \hspace{0.5cm}
     m_k^y=\frac{1}{L^2}\sum_{i,j}S^y(i,j,k)
 \end{equation*}
 where $S^x(i,j,k)$ and $S^y(i,j,k)$ denote the Cartesian components of the spin vector on the site $(i,j)$ belonging to the $k$-th sublattice and $L^2$ is the total number of lattice sites in each sublattice. The total magnetisation of the system is obtained by summing the contributions from all sublattices:

 \begin{equation*}
     M=\sqrt{M_x^2+M_y^2}
 \end{equation*}

 \noindent where,
  \begin{equation*}
{M_x = \sum_k m_k^x \hspace{0.5cm} M_y = \sum_k m_k^y}
 \end{equation*}
 
 We compute corresponding magnetic susceptibility by 
 \begin{equation*}
     \chi=\frac{3L^2}{k_BT}(\langle M^2 \rangle -\langle M \rangle ^2)
 \end{equation*}
 where $T$ is the temperature and $k_B$ is Boltzman constant. The specific heat $C$ is defined as , 
 \begin{equation*}
     C=\frac{d\langle E \rangle}{dT}
 \end{equation*}
 where $\langle E \rangle $ is the average energy density calculated from the Hamiltonian. 
 
\vspace{0.2cm}
\section{Results and Discussion}
\subsection{Isotropic regime:$D=0$}

\begin{figure}[h]
  \centering
  (a)\includegraphics[width=0.5\columnwidth]{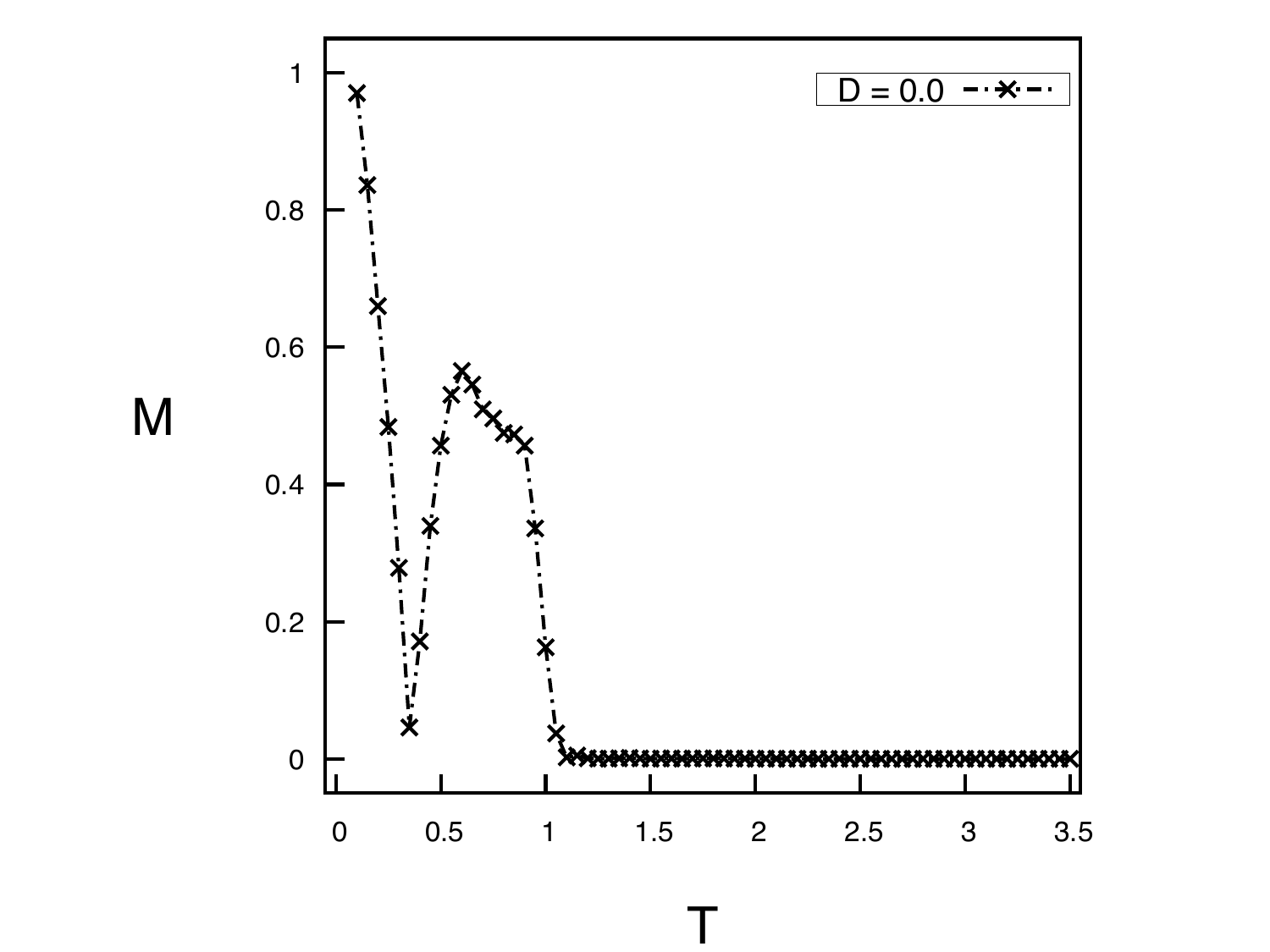}
  (b)\includegraphics[width=0.5\columnwidth]{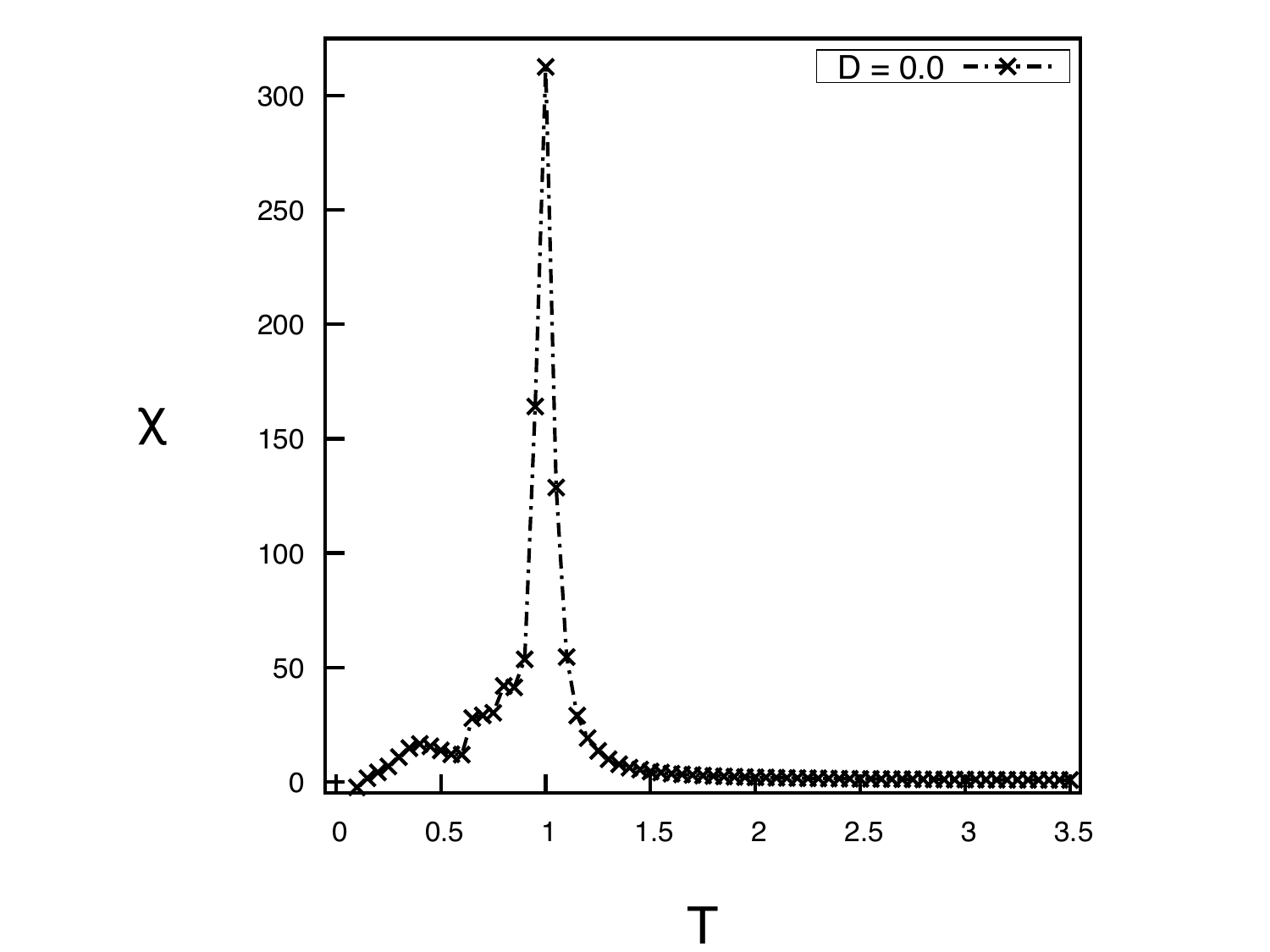}
  (c)\includegraphics[width=0.5\columnwidth]{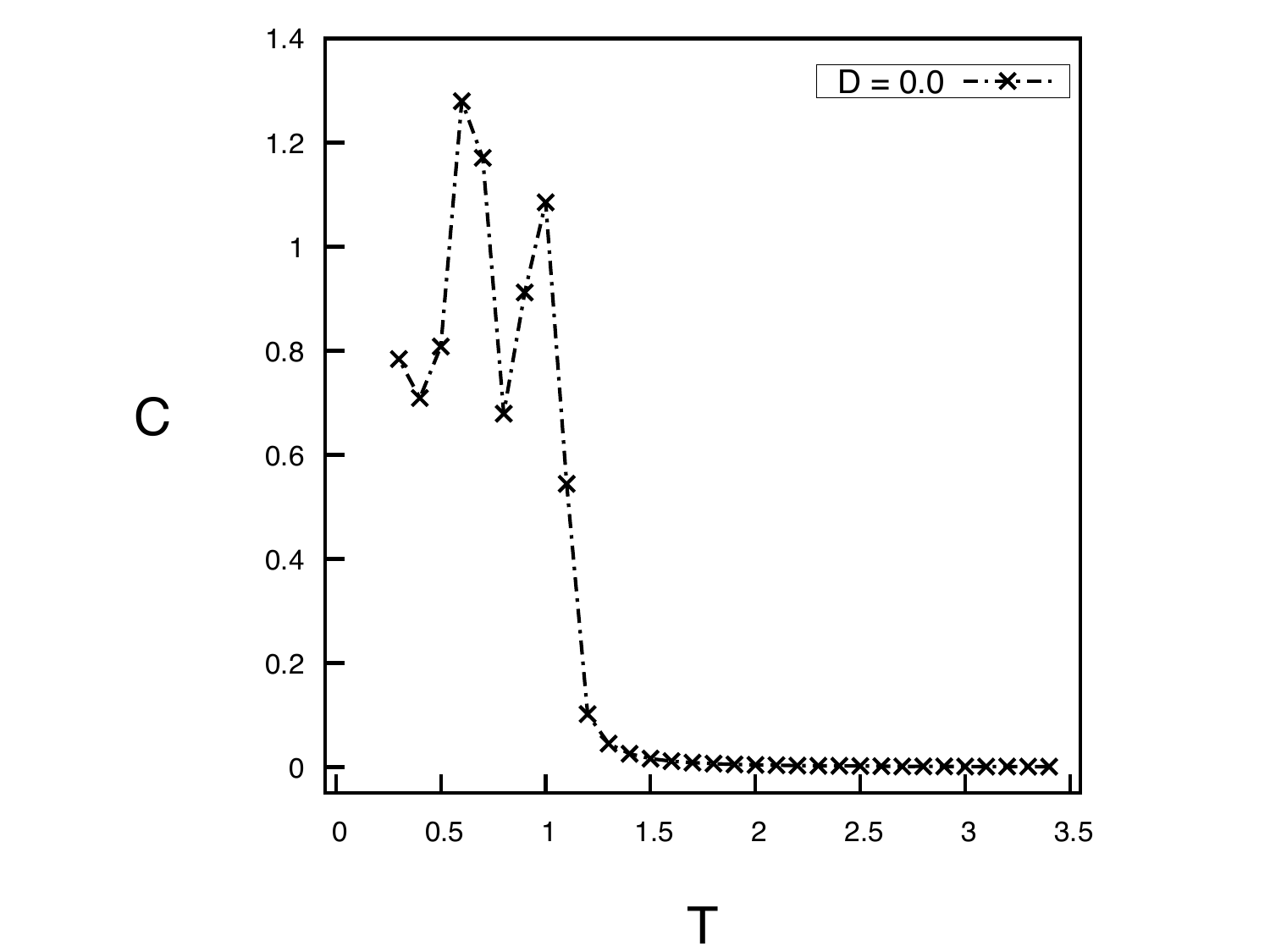}
  \caption{Thermal variation of (a) total magnetisation $(M)$ , (b) susceptibility $(\chi)$ and (c) specific heat $(C)$ for isotropic regime $(D=0)$. The transition temperature is found to be \textcolor{red}{ $T_c \approx 1.0$} and the compensation temperature \textcolor{red}{$T_{comp} \approx 0.40$}, both determined from the  positions of the peaks of the susceptibility. The corresponding specific heat $C(T)$ also exhibits peaks at the respective temperatures.}
  \label{fig:D=0}
\end{figure}

First, we discuss the thermodynamic phase transition and compensation phenomenon of a trilayered 6-state clock model at zero anisotropy($D=0$). All six clock states are energetically favourable in this regime, as there is no directional bias. The intralayer ferromagnetic coupling is taken for A-type atoms as  $J_{ferro}^{AA}= 0.2$ and for B-type atoms as $J_{ferro}^{BB}= 1.0$. The interlayer antiferromagnetic coupling strength is taken as $J_{antiferro}^{AB}=-0.1$. Initially, the system is at high-temperature disordered paramagnetic state, characterised by random spin orientations and zero magnetisation. The system is slowly cooled down with small temperature step $dT=0.05$. As the temperature decreases, magnetic ordering sets in. Figure-\ref{fig:D=0}(a) shows the growth of total magnetisation $M(T)$, which increases continuously from zero to a non-zero finite value at a well-defined critical temperature $T_c$. The transition temperature $T_c$ is precisely identified from the position of the peak of the magnetic susceptibility $\chi(T)$ as shown in Figure-\ref {fig:D=0}(b).

Interestingly, below $T_c$, the total magnetisation again vanishes at a lower temperature $T_{comp}$, referred to as \textit{compensation temperature}. This occurs even though the sublattices (individual layers) remain {\it partially} ordered. Compensation is a peculiar balancing situation where the total magnetisation of all layers vanishes, despite the nonzero sublattice ordering. The susceptibility curve $\chi(T)$ exhibits a secondary peak at this temperature, further supporting the presence of a sharp magnetic crossover. The corresponding specific heat $C(T)$ shown in Figure-\ref{fig:D=0}(c), also reflects these thermodynamic features with distinct peaks at both $T_c$ and $T_{comp}$.
The compensation phenomenon originates from the competing magnetic interactions: each layer offers ferromagnetic intra-layer coupling, favouring uniform spin alignment within the layer, where adjacent layers are coupled via anti-ferromagnetic interlayer interactions leading to repel layer magnetisation. 

\begin{figure}[h]
    \centering
   (a) \includegraphics[width=0.45\linewidth]{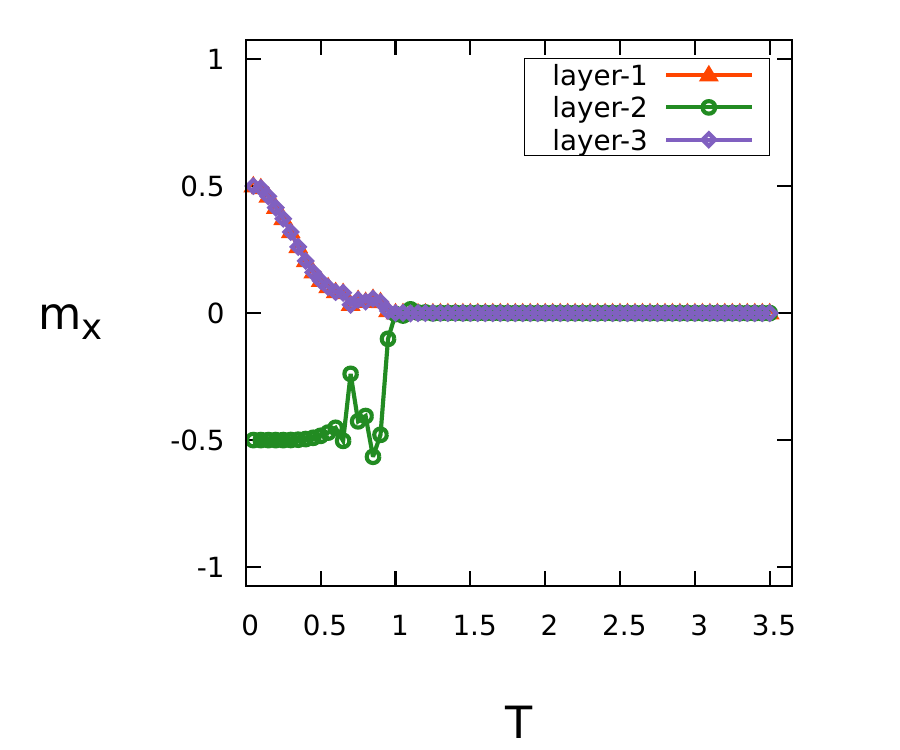}
    (b) \includegraphics[width=0.45\linewidth]{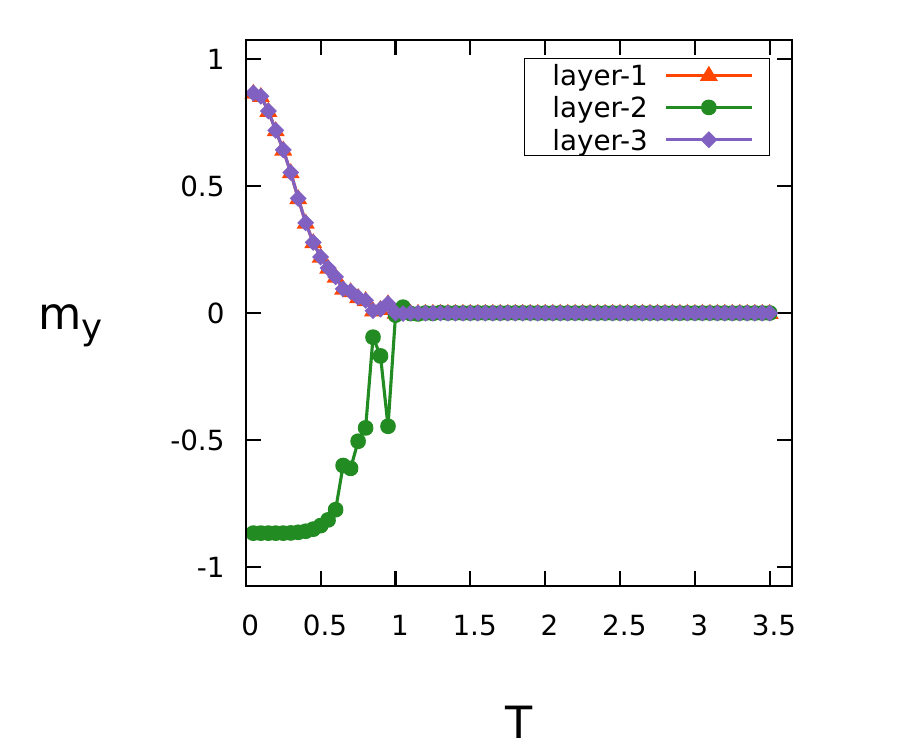}
    \caption{Thermal variation of sublattice magnetisation component (a) $m_x$ and (b) $m_y$ for three layers in the isotropic regime $D=0$. Layers 1 and 3 exhibit positive $m_x$ and $m_y$ values, while layer 2 shows negative contributions. At compensation temperature $T_{comp}=0.40$, each sublattice retains a finite magnetisation, but their vector sum cancels out resulting in vanishing total magnetisation.}
    \label{fig:sublattice-x-D0}
\end{figure}

To acquire a more nuanced perspective on the magnetic ordering in each layer, we analyse the sublattice magnetisation of the individual layers across the temperature. Figure-\ref{fig:sublattice-x-D0} shows the sublattice magnetisation components $m_x$ and $m_y$ as functions of temperature for three distinct layers. These profiles describe the growth of magnetic ordering in each layer. At $D=0$, where no easy-axis anisotropy is present, the system does not energetically favour in any specific direction in the $xy$ plane. Consequently, both $m_x$ and $m_y$ develop finite values below the transition temperature, indicating spontaneous symmetry breaking. As shown in Figure-\ref{fig:sublattice-x-D0}(a) or (b) all three layers exhibit ferromagnetic ordering below the transition temperature $T_c$ with each layer developing finite magnetisation. However, due to the antiferromagnetic interlayer coupling, the magnetisations of adjacent layers are oriented in opposite directions. In particular, the middle layer tends to be antiparallel to the outer layers, leading to cancellation of total magnetisation at $T_{comp}$.

\begin{figure}[h]
  \centering
  \includegraphics[width=0.9\textwidth]{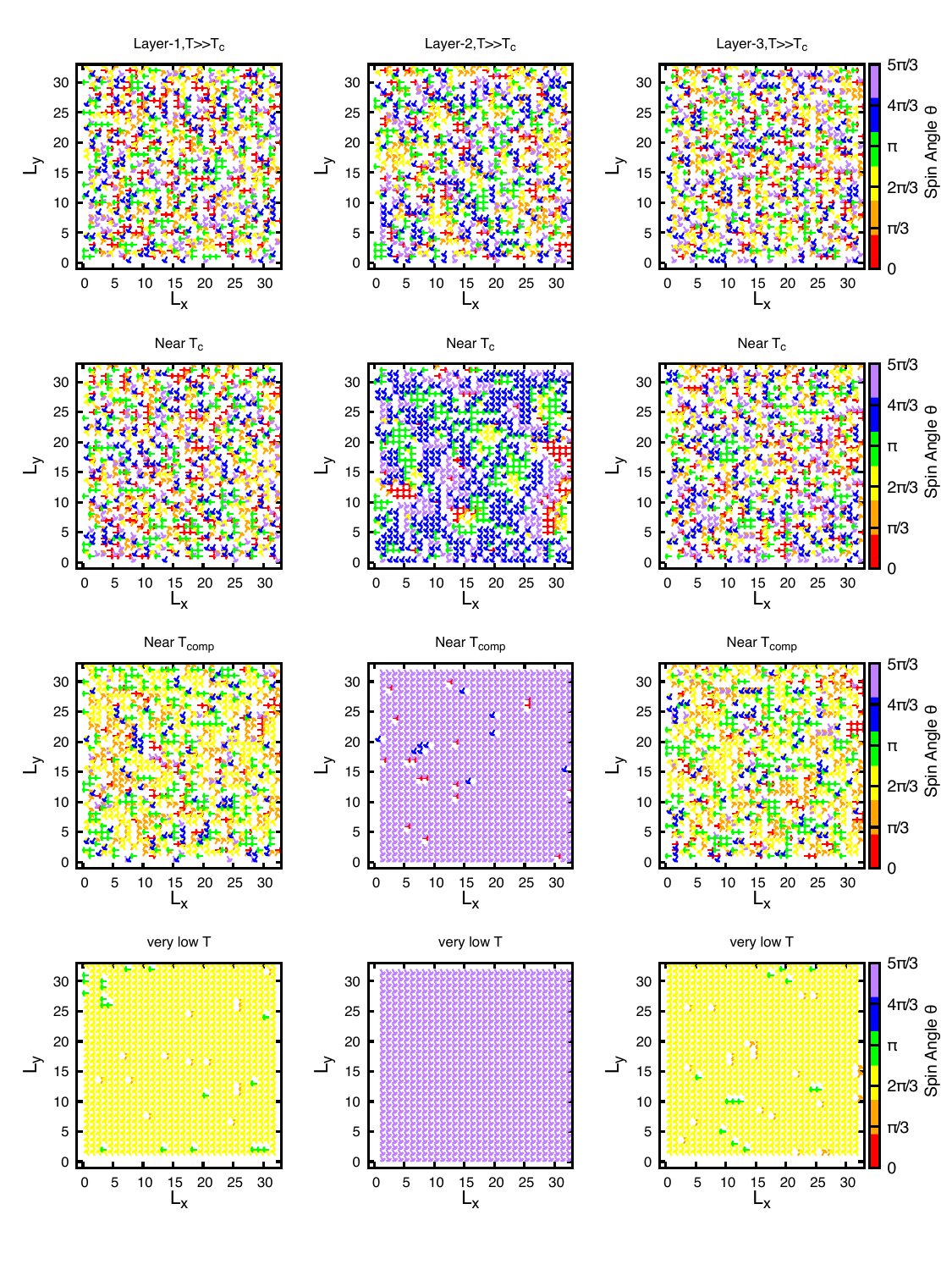}
  \caption{Spin morphology for layers at  four characteristic temperatures ($T_{high}$, Near $T_c$, $T_{comp}$ and $T_{low}$) in the isotropic regime ($D=0$) for system size $L=32$. Each colour represents one of the six clock states corresponding to the spin angle. The snapshot illustrates the evolution from high-temperature paramagnetic phase through thermodynamic phase transition near $T_c$, the compensation point $T_{comp}$ to the low temperature ordered state.}
  \label{fig:D=0:spin_morphology}
\end{figure}

\begin{figure}
    \centering
    \includegraphics[width=0.5\linewidth]{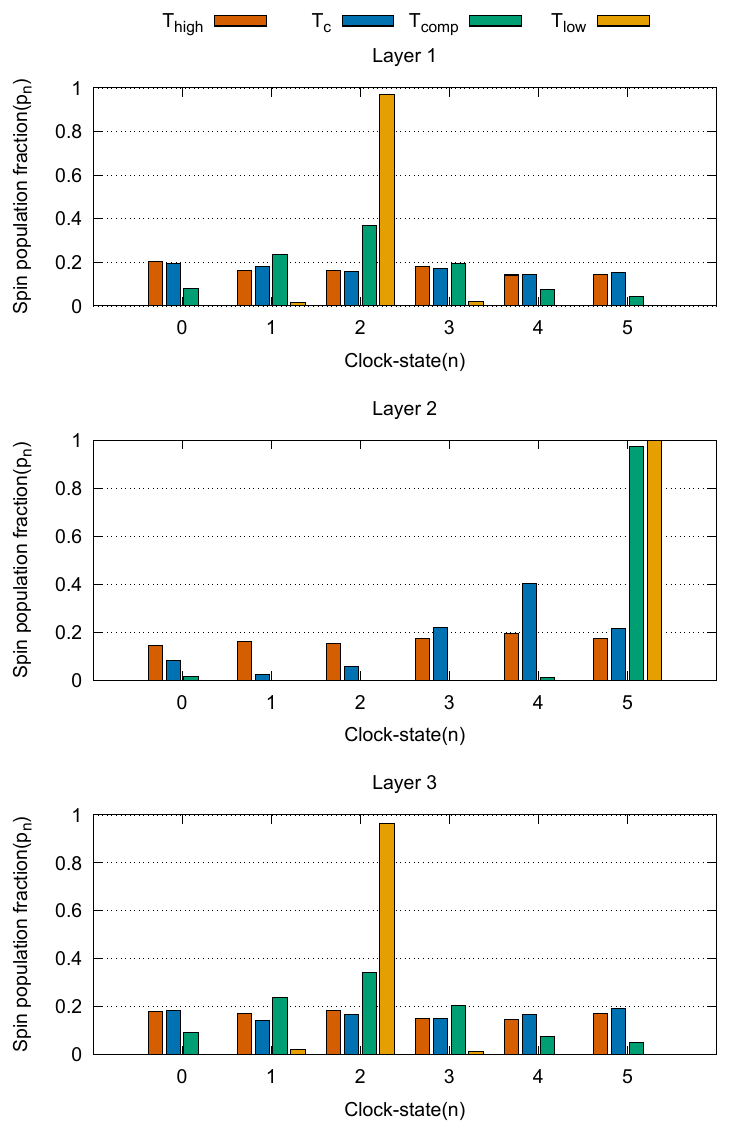}
    \caption{Normalised spin population as a function of six clock state (discrete spin angles) for three layers at four characteristic temperatures $T_{high}$, $T_c$, $T_{comp}$ and $T_{low}$. The vertical axis represents the fraction of spins in each angle state (normalized to the total spin count of 1024 in each layer). At high temperature, the distribution is equal, whereas upon cooling, distinct states(angles) eventually catch on: spins at layer 1 and 3 align at $\theta=2\pi/3$($n_i=2$), while layer 2 aligns at $\theta=5\pi/3(n_i=5)$, consistent with spin morphology. }
    \label{fig:spin-count-D0}
\end{figure}

To visualise the spin ordering, we also examine the lattice morphology at representative temperatures. Spin configurations snapshots displayed in Figure-\ref{fig:D=0:spin_morphology} reveal a transition from a disordered paramagnetic state at high temperature to a well-aligned domain at low temperature for different layers. At very high temperature $T>> T_c$ spin are randomly oriented in all three layers. Six colours denote the six specific clock states. For the isotropic regime $(D=0)$, at high temperature, all six directions are equally preferable. As temperature decreases, the spins at each layer progressively align, leading to the nucleation and growth of the magnetic domain, which marks the onset of long-range order and results a non-zero total magnetisation. Near $T_{comp}$, distinct domain patterns emerge with opposing magnetisation contributions, reflecting the layer-wise compensation behaviour. To provide more systematic and statistically quantiative analysis of the spin orientations
near the compensation temperature, the statistical distributions of the angles (orientations) of the spins in different layers
are studied and represented by bar graphs in Figure-\ref{fig:spin-count-D0}, through a comprehensive manner. The normalised spin population (fraction of total spins) is plotted as a function of six discrete clock states ( corresponding spin angle $0$, $\pi/3$, $2\pi/3$, $3\pi/3$, $4\pi/3$, $5\pi/3$) for each layer. At high temperature, the spin population is nearly uniform across all angles in all three layers, reflecting an isotropic regime where all spin states are equally probable. As temperature decreases towards the critical value, certain spin angles become more populated, indicating the onset of preferred alignment. Near compensation temperature, layer 1 and 3 exhibit a dominant population at $\theta=2\pi/3(n_i=2)$, whereas layer 2 prefers $\theta=5\pi/3(n_i=5)$.  Importantly, the other spin angles are populated such that net magnetisation of the whole system becomes zero, while each layer individually retains a finite magnetisation. Hence, compensation occurs. Upon further cooling to low temperature, layers 1 and 3 show nearly complete alignment at $\theta=2\pi/3$ while layer 2 remains oriented in $5\pi/3$, resulting nonvanishing total magnetisation. The visual and quantitative results together confirm the layered character of magnetic ordering and the competing interactions that drive the compensation phenomenon. \textcolor{blue}{This statistically quantitative representation of the spin orientations is consistent with that represented by qualitative morphological respresentation shown in Figure-\ref{fig:D=0:spin_morphology}, in the context of compensation.}

\subsection{Anisotropic regime: $D>0$}
\begin{figure}[h]
  \centering
  (a)\includegraphics[width=0.5\textwidth]{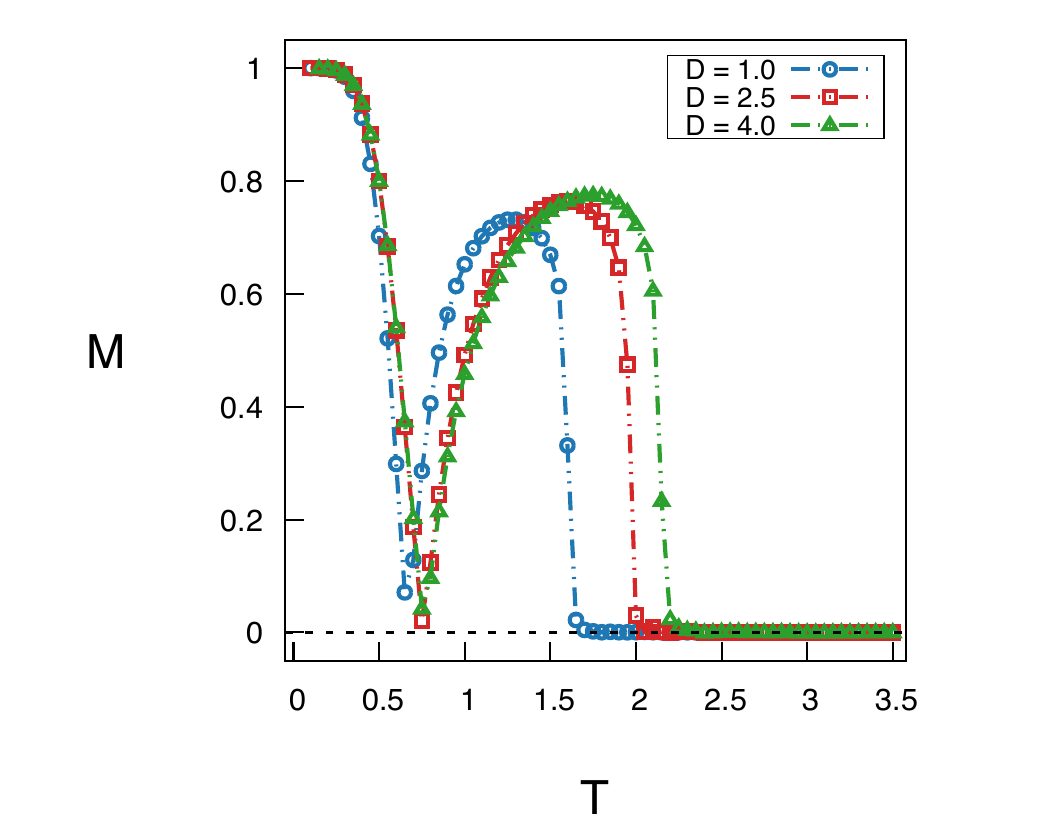}
  (b)\includegraphics[width=0.5\textwidth]{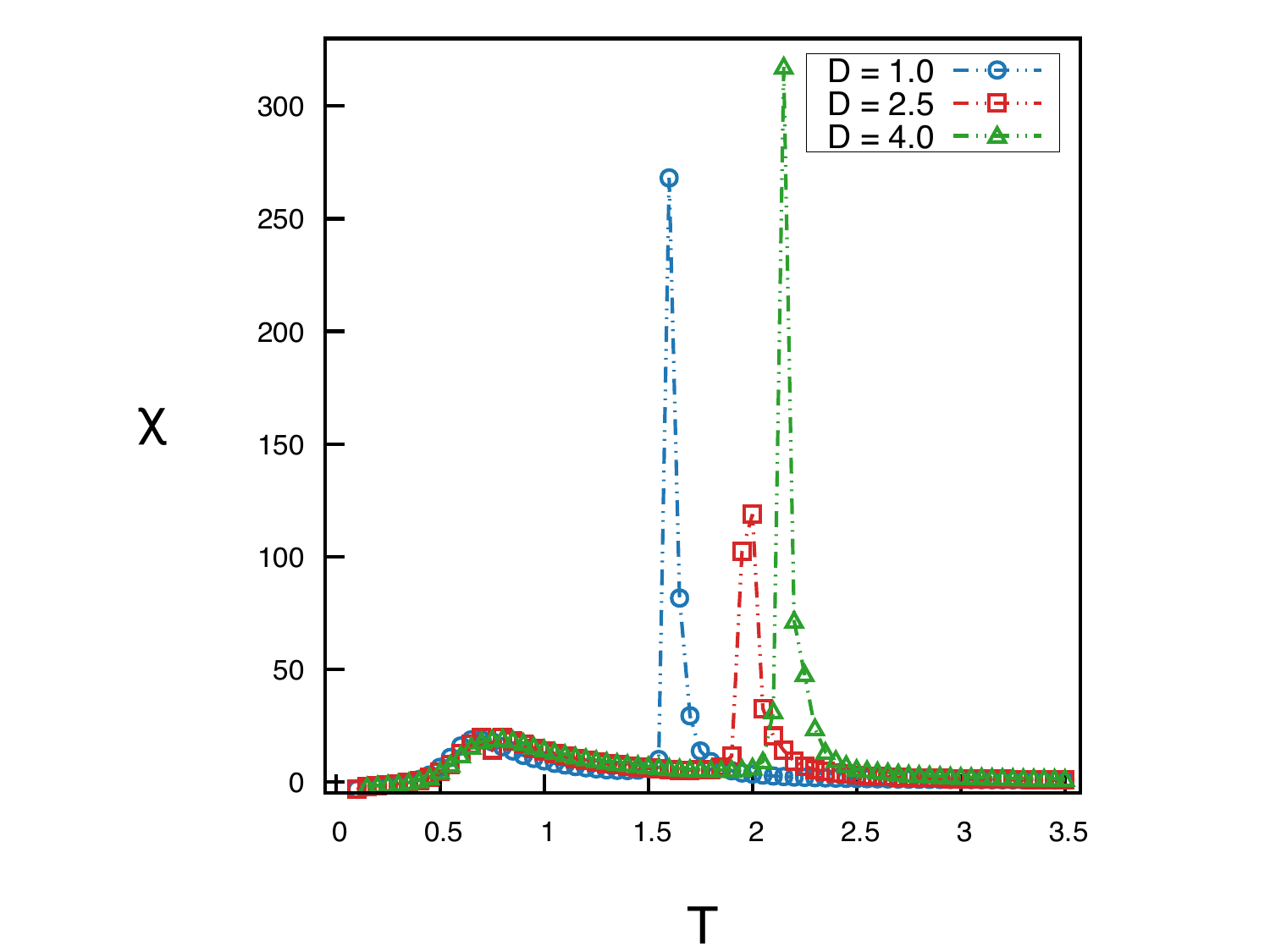}
  (c)\includegraphics[width=0.5\textwidth]{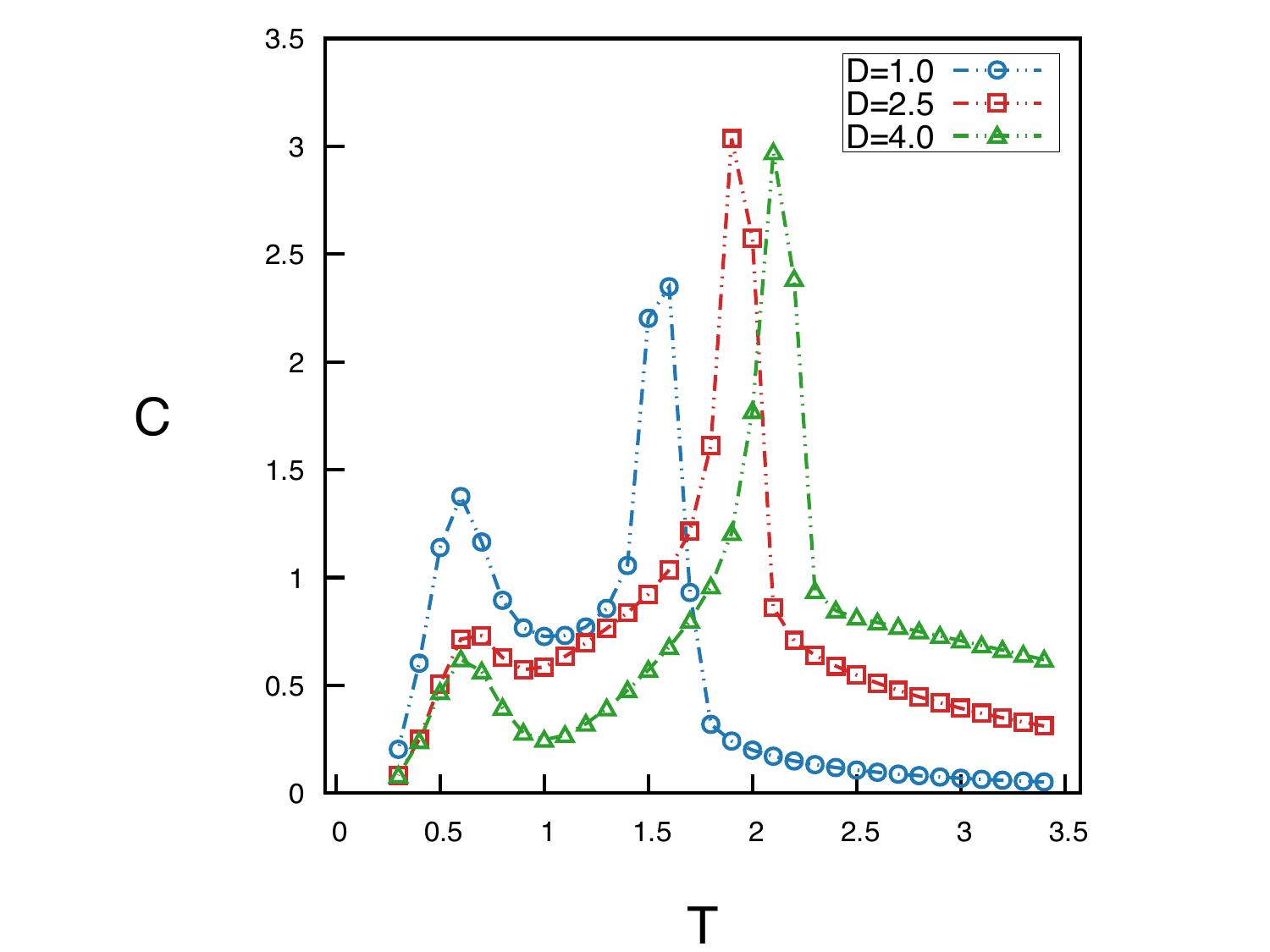}
  \caption{Thermal variation of (a) total magnetisation $(M)$, the dotted horizontal line is a reference for $M=0$, (b) susceptibility $(\chi)$ and (c) specific heat $(C)$ for different anisotropy values $D=1.0,2.5,4.0$. The critical temperature $T_c$ and compensation temperature $T_{comp}$ determined from the positions of susceptibility peaks, both shift towards higher values with increasing D, indicating enhanced thermal stability.}
  \label{fig:diff_D_observable}
\end{figure}

In this section, we explore the response of clock spins to the single site anisotropy. By varying the anisotropy parameter D, we systematically investigate the thermodynamics of the key observables such as magnetisation, susceptibility and specific heat. Figure-\ref{fig:diff_D_observable}(a) shows the thermal variation of total magnetisation as function of temperature for different anisotropy values. As temperature decreases, magnetisation grows, exhibiting a clear phase transition at $T_c$. Below $T_c$, the total magnetisation again vanishes at a lower temperature, denoted as the compensation point $T_{comp}$, indicating the compensation behaviour as already observed in the isotropic case. The critical temperature $T_c$ and compensation temperature $T_{comp}$ are precisely obtained from the peak positions of magnetic susceptibility curves Figure-\ref{fig:diff_D_observable}(b). The corresponding specific heat profiles further corroborate these findings by exhibiting peaks near both temperatures.

\begin{figure}
\centering
    \includegraphics[width=0.5\linewidth]{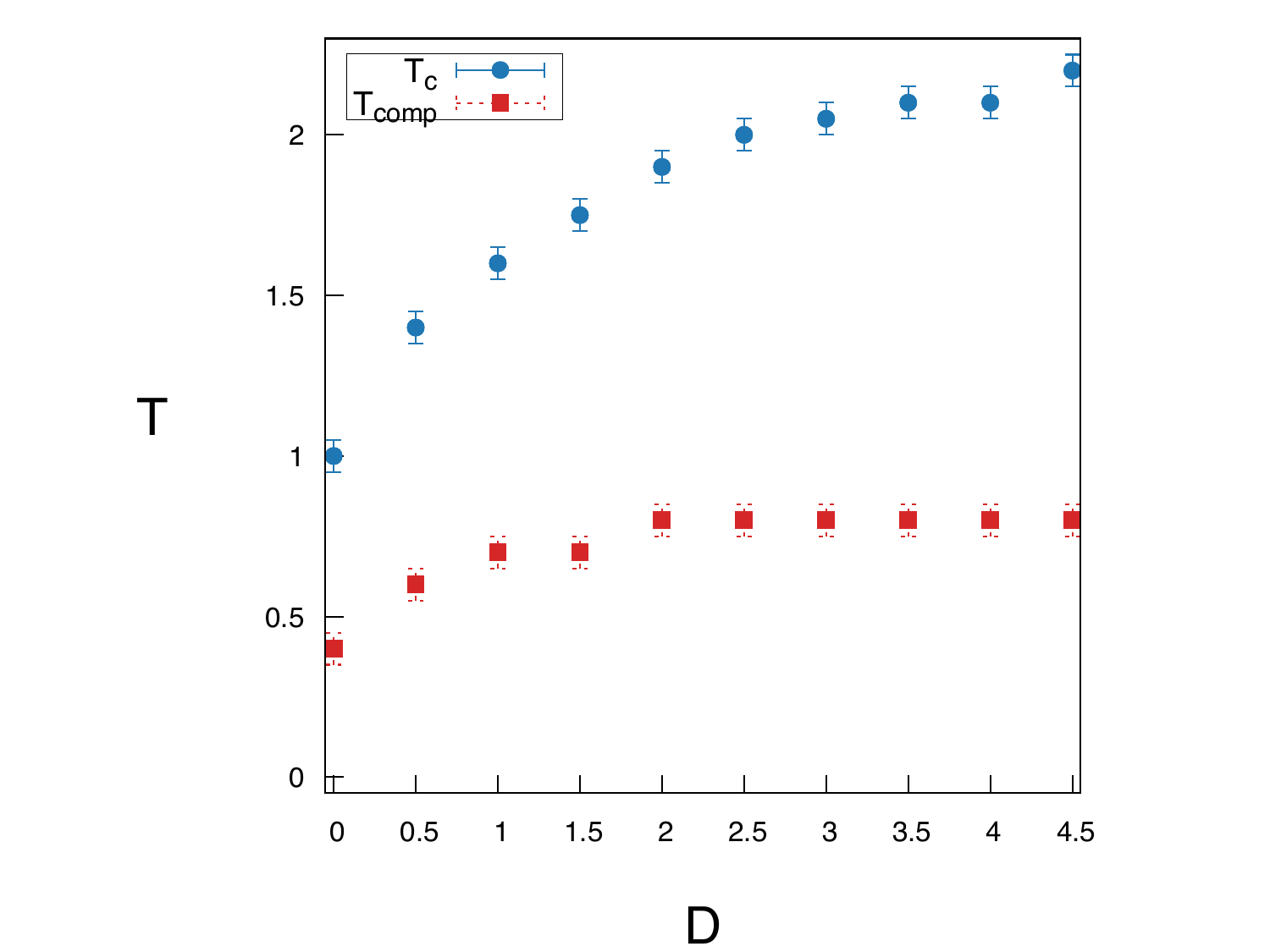}
    \caption{The critical temperature ($T_c$) and compensation temperature ($T_{comp}$) is plotted as a function of anisotropy strength ($D$).}
    \label{fig:phase}
\end{figure}

%%%Phase diagram %%%%%%%%%%%%
Now, if we increase anisotropy D, the position of the maximum of the susceptibility shifts toward the right (higher temperature), indicating the critical temperature $T_c$ increases as $D$ increases. The compensation point $T_{comp}$  also shifts slightly. This is captured in the phase diagram in $D-T$ plane shown in Figure-\ref{fig:phase}. This diagram summarises the evolution of both the thermodynamic phase transition temperature $(T_c)$ and the compensation temperature $(T_{comp})$ as a function of anisotropy $D$. A monotonic increase in $T_c$ is observed with increasing $D$, indicating that a stronger single-site anisotropy stabilizes the ordered phase and shifts the thermodynamic transition to higher temperatures. In contrast, $T_{comp}$ shows only a slight upward trend with increasing $D$ and remains well separated from $T_c$ across the probed range. The critical temperature $T_c$ is prescribed by collective critical fluctuations, which are highly sensitive to the anisotropy parameter $D$. Increasing $D$ enhances spin alignment and restrains thermal disorder, thereby stabilizes the ordered phase and shifts $T_c$ to higher temperatures. Besides, the compensation temperature $T_{comp}$ does not correspond to bonafide thermodynamic singularity. Instead, it arises from a subtle gross cancellation among the magnetisations of individual sublattices. This balance depends primarily on competing the intra and inter-layer exchange interactions and local spin configurations rather than on long-range critical fluctuations, resulting in only a weak variation of $T_{comp}$ with D.

\begin{figure}
  \centering
  (a)\includegraphics[width=0.5\textwidth]{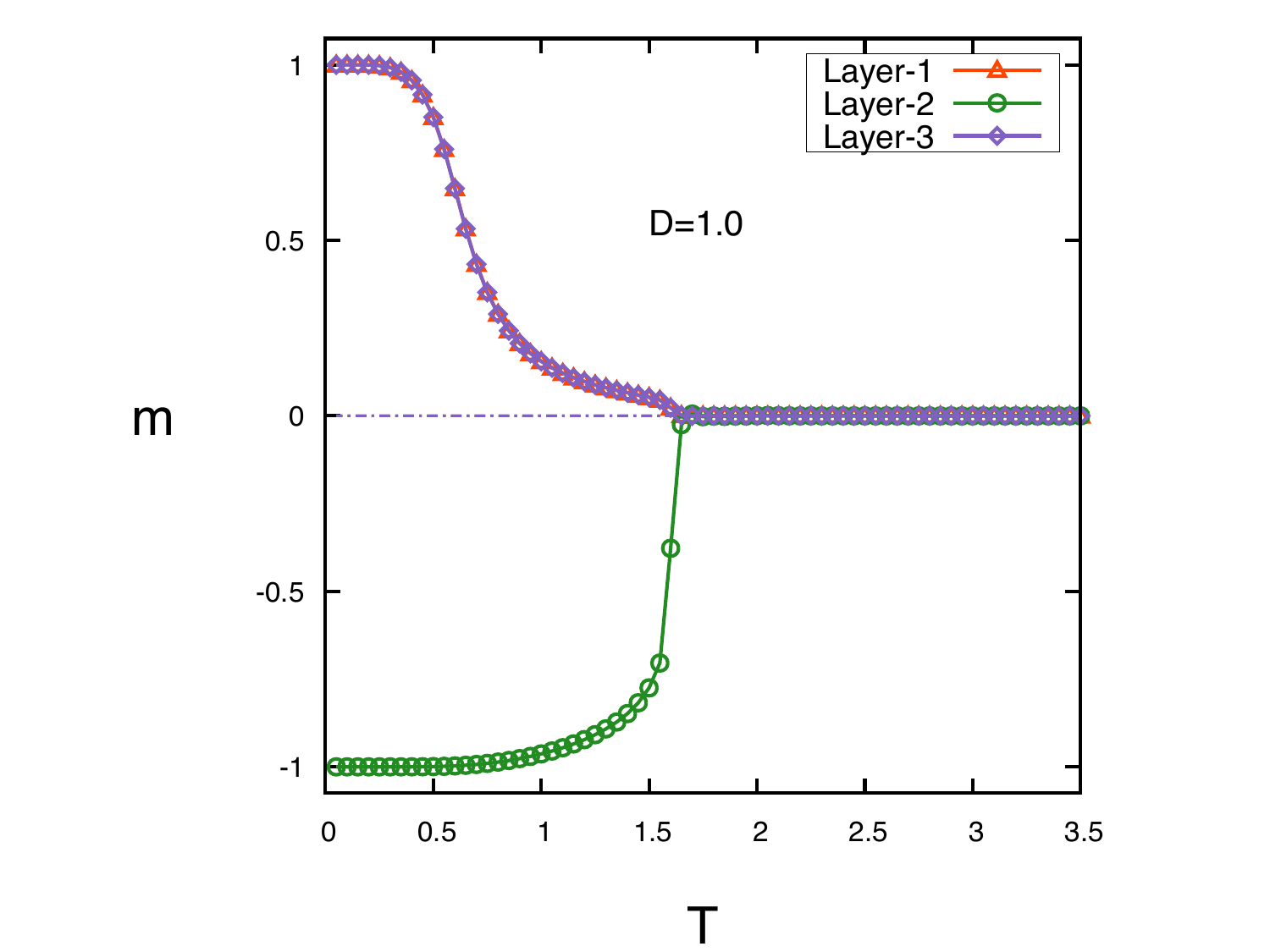}
  (b)\includegraphics[width=0.5\textwidth]{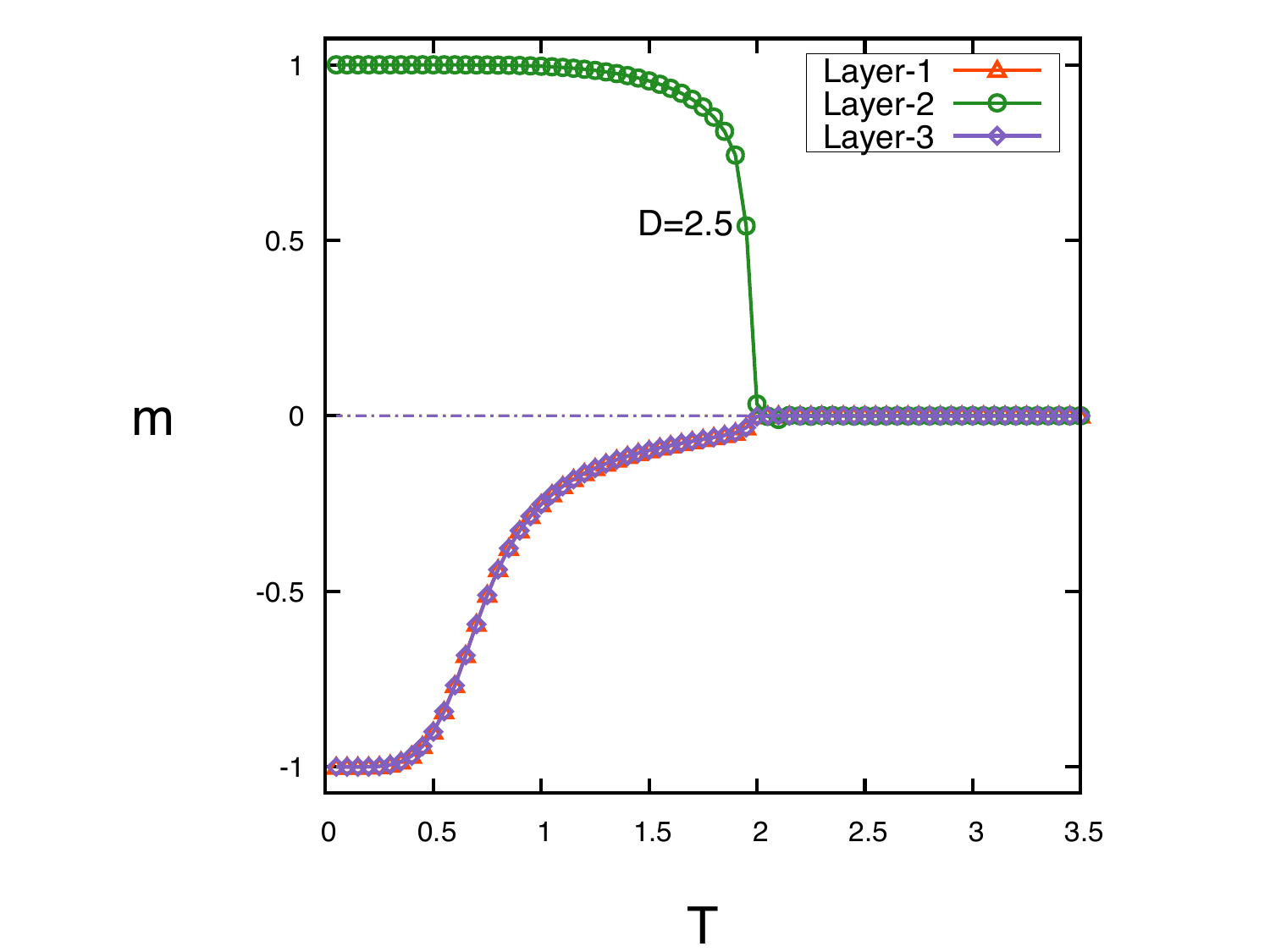}
  (c)\includegraphics[width=0.5\textwidth]{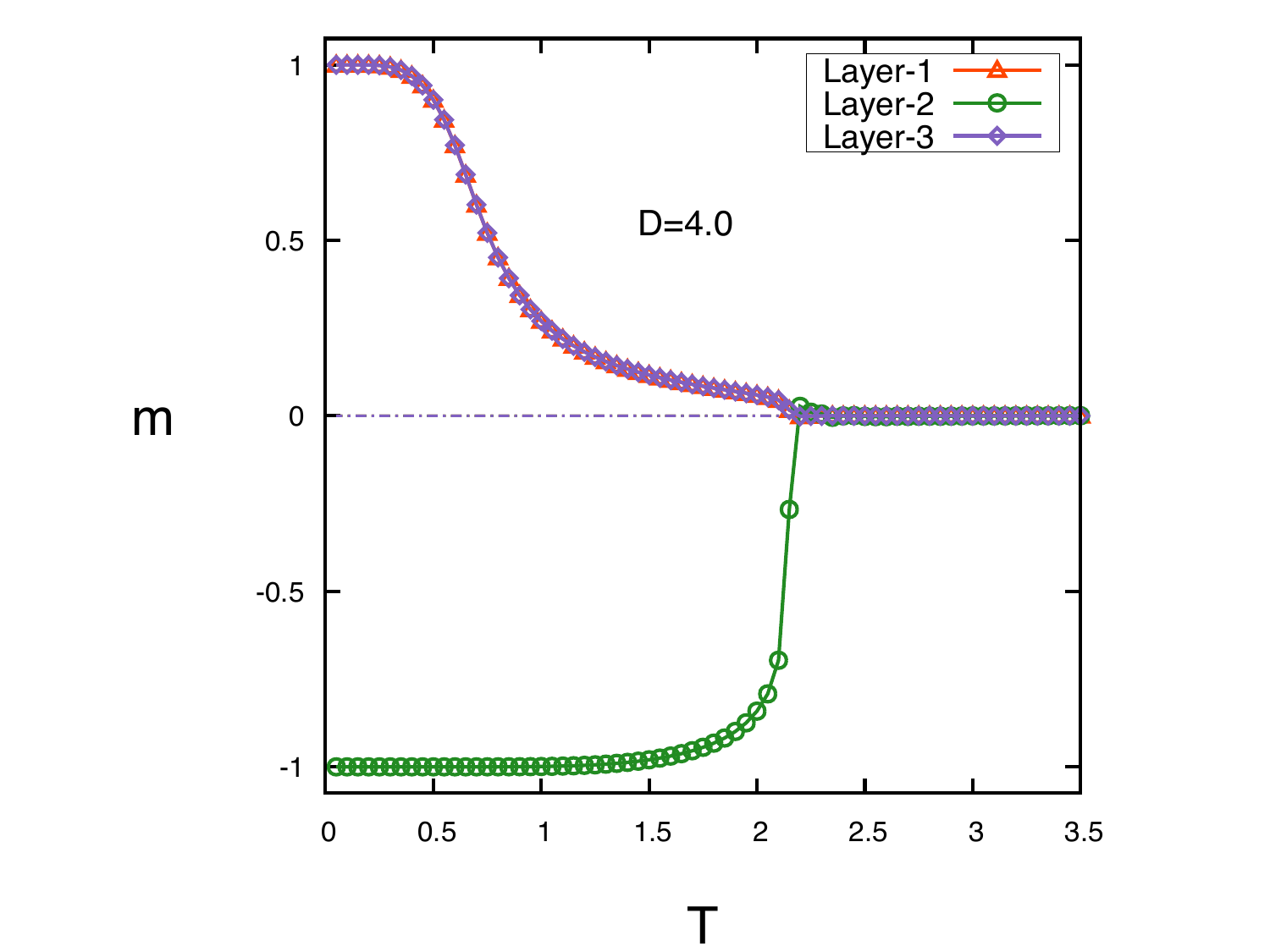}
  \caption{Sublattice magnetisation componets $m_x$(solid line) and $m_y$(dashed line)for three different anisotropy $D$ values. In all cases $m_x$ grows to finite values while $m_y$ remains zero across the temperature range. Layers 1 and 3 show magnetization opposite to that of layer 2. A clear shift of the compensation temperature $T_{comp}$ toward higher values is observed with increasing $D$.}
  \label{fig:diff_D_sublattice}
\end{figure}

We also analyse the sublattice magnetisation of different layers for different strengths of anisotropy (Figure-\ref{fig:diff_D_sublattice}). As positive $D$ influences magnetic ordering along $x$ direction, $y$ component of magnetisation $m_y$ remains zero always. As we decrease the temperature, $m_x$ grows in each of the layers, resulting in a thermal phase transition of the total system at a certain temperature $T_c$. This is shown in Figure-\ref{fig:diff_D_sublattice}. Due to the antiferromagnetic coupling between the adjacent layers, magnetic orderings are opposite in the respective layers. At a certain temperature ($T_{comp}$), they cancel each other, leading to a net magnetisation of zero (even with nonzero sublattice magnetisation).  The compensation temperature $T_{comp}$ also increases (with small variation)  as the strength of anisotropy increases.

\begin{figure*}
  \centering
  \includegraphics[width=0.9\textwidth]{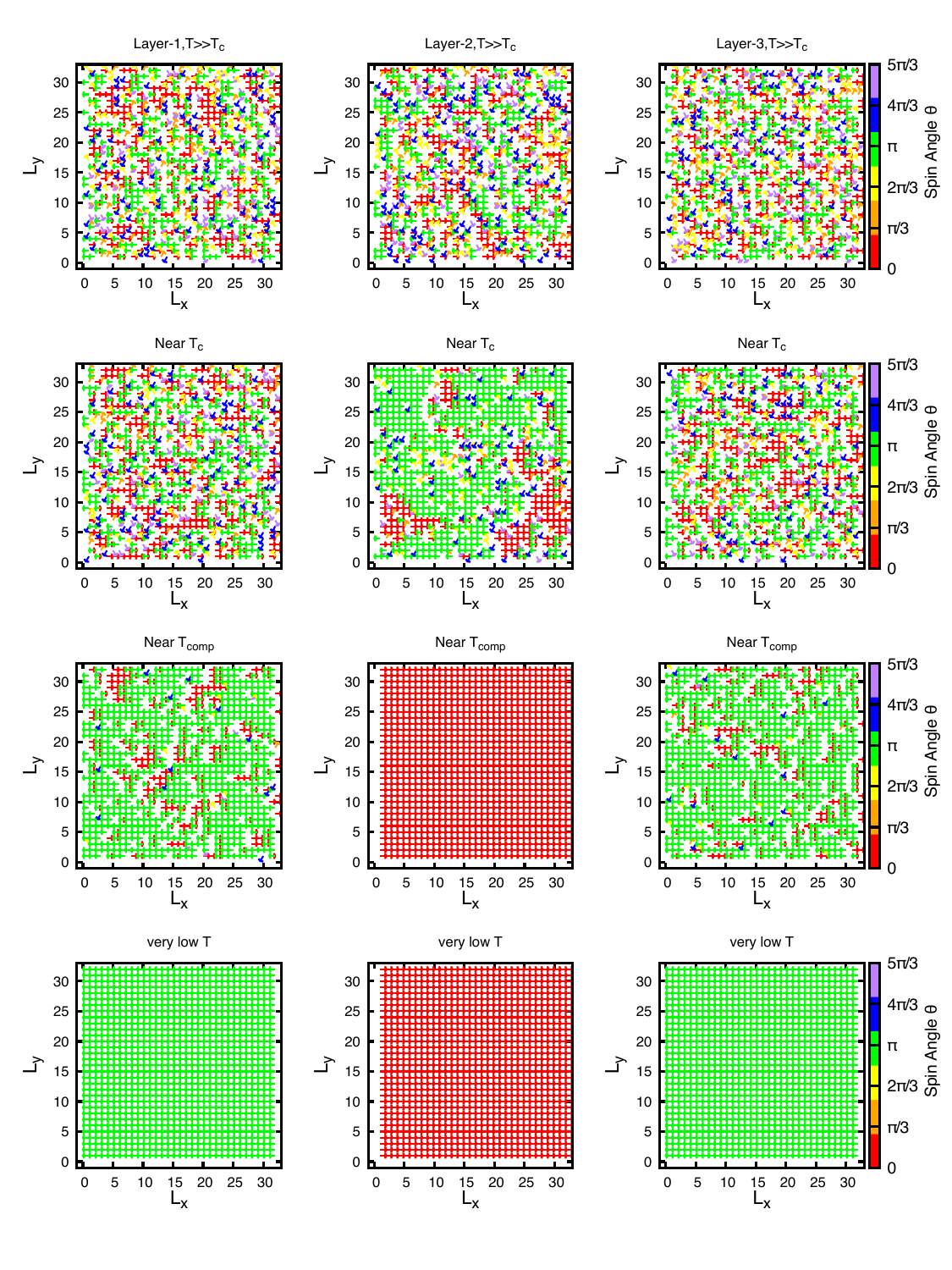}
  \caption{Spin morphology for layers at  four characteristic temperatures ($T_{high}$, Near $T_c$, $T_{comp}$ and $T_{low}$) in the anisotropic regime ($D=2.0$) for system size $L=32$. Each colour represents one of the six clock states corresponding to the spin angle. The snapshot illustrates the evolution from high-temperature paramagnetic phase through thermodynamic phase transition near $T_c$, the compensation point $T_{comp}$ to the low temperature ordered state.}
  \label{fig:diff_D_spin_morphology}
\end{figure*}

 The spin morphology at different temperatures for anisotropy $D=2.0$ is shown in Figure-\ref{fig:diff_D_spin_morphology}. At high temperature $T>>T_c$ the spins
are randomly oriented at all layers, resulting in zero magnetisation. Due to the positive anisotropy majority of spins are aligned along $0$ or $\pi$. Near the critical temperature $T_c$, thermal fluctuations are still strong enough to disrupt complete spin alignment, but a tendency emerges for spins to align along either the positive or negative $x$ direction. This marks the onset of magnetic domain growth, where regions of correlated spins begin to form and compete. At the compensation point $T_{comp}$, the system exhibits a striking balance: the spins in the two outer layers align in one direction, while those in the middle layer orient in the opposite direction. As a result, the net magnetisation of the entire system cancels out to zero, even though each layer retains a finite magnetisation. This phenomenon is characteristic of ferrimagnetic-like ordering, where sublattice magnetisations compensate each other. At very low temperatures,  the system reaches a fully ordered state. In this regime, the spins in the two outer layers are aligned along the negative $x$-direction, while the spins in the middle layer are aligned along the positive $x$-direction. This configuration reflects the combined effects of interlayer exchange interactions and anisotropy.

\begin{figure}
    \centering
    \includegraphics[width=0.5\linewidth]{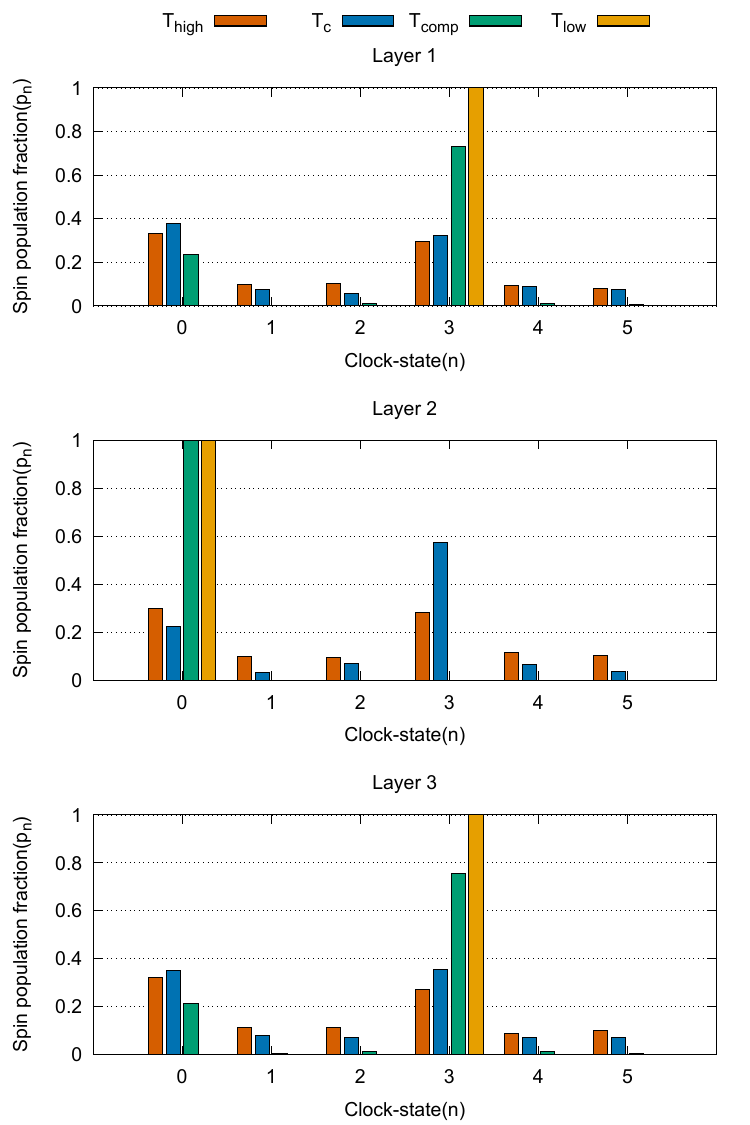}
    \caption{Normalised spin population as a function of six clock state (discrete spin angles) for three layers for the anisotropic regime ($D=2.0$) at four characteristic temperatures $T_{high}$, $T_c$, $T_{comp}$ and $T_{low}$. The vertical axis represents the fraction of spins in each angle state (normalised to the total spin count of 1024 in each layer). A positive anisotropy favours spin alignment along $\theta=0(n_i=0)$ or $\theta=\pi(n_i=3)$ at high temperature majority of spins predominantly occupy either of these two directions. As the temperature decreases, layers 1 and 3 progressively shift their population toward $\theta=\pi(n_i=3)$, while layer 2 spin is aligned at $\theta=0(n_i=0)$  near the compensation point, consistent with spin morphology. }
    \label{fig:spin-count-D2}
\end{figure}

 To further elucidate the role of anisotropy in shaping the spin morphology, we also examined the angular distribution of spins for the case of $D=2.0$, where the easy-axis energetically favours spin orientations along $\theta=0$ or $\theta=\pi$. This is shown in Figure-\ref{fig:spin-count-D2}. At high temperature, the majority of spins in all layers are distributed near $\theta=0(n_i=0)$, and $\theta=\pi(n_i=3)$ reflecting the dominant easy-axis alignment imposed by the positive anisotropy. Upon cooling, the distribution becomes progressively more ordered, with the relative populations $\theta=0$ and $\theta=\pi$ starting to compete. Near the compensation temperature, a distinct redistribution occurs: layers 1 and  3 exhibit a stronger spin population at $\theta=\pi(n_i=3)$, whereas layer 2 retains it alignment at $\theta=0(n_i=0)$. This antiparallel arrangement of sublattices leads to a vanishing total magnetisation while preserving finite sublattice magnetisation. At very low temperature, the system reaches a nearly complete alignment where layers 1 and 3 are oriented along $\theta=\pi$ and layer 2 aligned along $\theta=0$. At sufficiently low temperatures, the system recovers a finite total magnetisation as thermal fluctuations subside. Thus, while both the isotropic ($D=0$) and anisotropic ($D=2.0$) systems exhibit a compensation behaviour governed by the redistribution of opposing spin populations, the positive anisotropy merely selects preferred axes ($\theta=0,\pi$) without changing the underlying mechanism.

\subsection{Finite size effects:}

\begin{figure}
  \centering
%  (a)\includegraphics[width=0.5\textwidth]{Ul-L.pdf}
  (a)\includegraphics[width=0.5\textwidth]{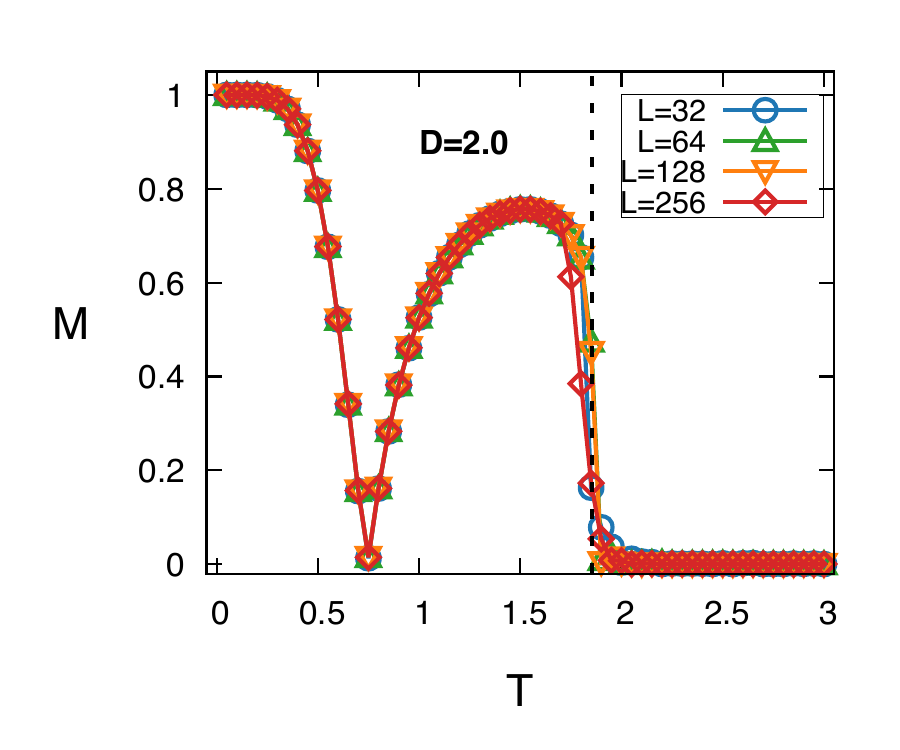}
  (b)\includegraphics[width=0.5\textwidth]{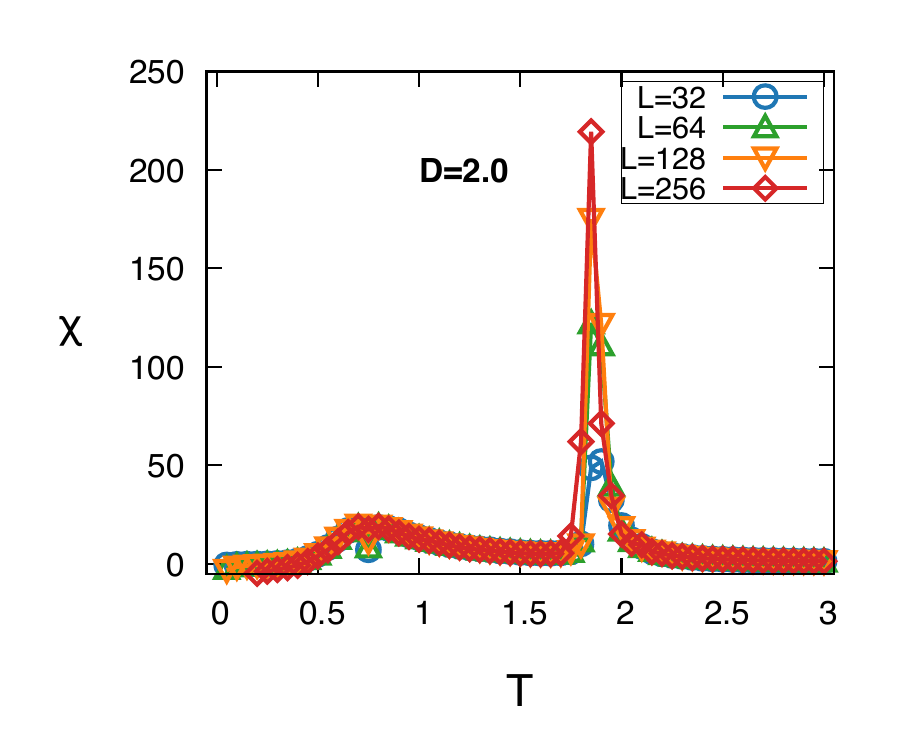}
  \caption{(a) Total magnetisation $M$ and (b) corresponding susceptibility $\chi$ is plotted against temperature $T$ for different system sizes $L=32,64,128,256$ for anisotropy $D=2.0$.}
  \label{fig:finite_size}
\end{figure}

%%%%%%Finite%size%%%Analysis%%%%%%%%%%%%%%%%%%%%%%%%%%%%%%%%%%%%%%%%%%%
Finite-size effects play a crucial role in accurately characterising the thermodynamic phase transitions in numerical simulations. To investigate the finite-size behaviour of the clock model in anisotropic regime ($D=2.0$), we performed simulations for different system sizes $L=32, 64, 128, 256$. The  magnetisation $M(T)$ and corresponding susceptibility $\chi(T)$ are studied as a function of temperature. The numerical results are shown in Figure-\ref{fig:finite_size} for the considered system sizes.  Figure-\ref{fig:finite_size}(a) shows the magnetisation $M(T)$, revealing the expected order-disorder behaviour as the temperature crosses the transition ($T_c$) and the low temperature compensation. The susceptibility curves for different system sizes (shown in Figure-\ref{fig:finite_size}(b))manifest two salient features: a prominent peak at the thermodynamic critical temperature and a secondary smeared peak at the compensation point.  The height of the peak of the susceptibility at $T_c$ exhibits a systematic increase with increasing system sizes, confirming the growth of critical correlation for the second-order phase transition. The susceptibility is believed to be divergent in the thermodynamic limit ($ L\rightarrow \infty$).  In contrast, the secondary smeared peak associated with the compensation point does not show such behaviour; its height remains nearly constant across the studied system sizes.  This suggests that such compensation is not a critical transition, but rather a finite, non-divergent reorganisation of the magnetisation in the magnetic sublattices. The finite-size analysis confirms that the clock model undergoes a thermodynamic phase transition at $T_c$
, where critical fluctuations become system-size dependent, while the compensation point reflects a non-critical feature, inherent to the system's response. 
\newpage

\section{Concluding remarks}

The magnetic compensation behaviour (vanishing of the total magnetisation for nonzero sublattice magnetisation) in a trilayered
ferrimagnetic system is an interesting effect studied in discrete symmetric spin models (e.g., Ising, Blume-Capel etc.). This
compensation effect has not been investigated before in any 
discrete symmetric planar magnetic models. The present study explores the compensation behaviours in discrete symmetric
planar magnetic model, i.e., clock model. We have chosen 6-state clock model (with anisotropy) in a trilayered structure. The ferromagnetic nearest neighbour intraplanar interaction and the antiferromagnetic interplanar nearest neighbour interactions are considered. The response of the system is investigated through the Monte Carlo methods using Metropolis single spin flip algorithm.

The sublattice magnetisations of three different layers, the total magnetisation, the susceptibility and the specific heat are
calculated. These quantities are studied as functions of the temperature in equilibrium. For the isotropic case ($D=0$), the 
low temperature compensation (vanishing of the total magnetisation for nonzero values of sublattice magnetisations) and 
high temperature critical 
(each sublattice magnetisation vanish) behaviour are identified. These two special points (critical and compensation) are 
identified as two special values of the temperatures which generally provide the maxima of susceptibility and the specific heat.

We have thoroughly investigated the effects of anisotropy (single-site). How do the critical and compensation temperatures
changes with the variation of anisotropy ? We have observed that both critical and compensation temperature increases with
the increase of anisotropy and finally saturate. The comprehensive phase diagram is provided. The morphologies of the spin
orientations (discrete) are investigated in four different regimes of the temperature. 

The finite size analysis is done through the temperature dependence of the susceptibility for variuos system sizes. The height of the peak of the susceptibility has been found to grow with the system size. \textcolor{blue}{To estimate the critical exponents and to determine the universality class, further rigourous
simulations are required, which is beyond the scope of the present study.}

The q-state clock model maps to continuous symmetric XY model in the limit $q \to \infty$. It would be interesting to 
study the compensation behaviours, in the q-state clock model, by varying $q$. The extrapolation in the limit $q \to \infty$
should reach the compensation temperature for XY model. Moreover, the compensation in disordered (may be modeled by randomly
distributed anisotropy or random field or both) system would really be an interesting and important investigation,
as the continuous symmetric planar ferromagnet (XY model) showed interesting behaviours in the case of 
randomly distributed anisotropy\cite{xy-rand} and random field\cite{rf-xy}. 

\begin{center}{----------------------------------------------}\end{center}

\vspace{0.2cm}
%\section{Discussions}
\vskip 1cm
\noindent {\bf Acknowledgements:}  
 We are thankful to the computational facilities provided by Presidency University, Kolkata. 
\vskip 0.2cm
%%%%%%%%%%%%%%%%%%%%%%%%%%%%%%%%%%%%%%%%%%%%%%%%%%%%%%%%%%%%%%%%%%%%%%%%%%%%%%%%%%%%
\noindent {\bf Data availability statement:} Data will be available on reasonable request to Olivia Mallick.

\vskip 0.2cm

\noindent {\bf Conflict of interest statement:} We declare that this manuscript is free from any conflict of interest.
\vskip 0.2cm

\noindent {\bf Funding statement:} No funding was received, particularly to support this work.

\vskip 0.2cm

\noindent {\bf Authors’ contributions:} Olivia Mallick developed the code, collected the data, prepared the figures, analysed the results and wrote the manuscript.
Muktish Acharyya conceptualised the problem, analysed the results and wrote the manuscript.

%\end{twocolumn}
 
 %%%%%%%%%%%%%%%%%%%%%%%Fig-1%%%%%%%%%%%%%%%%%%%%%%%%%%%%%%%%%%%
%%%%%%%%%%%%%%%%%%%%%%%%%%%%%%%%%%%%%%%%%%%%%%%%%%%%%%%%%%%%%%

\newpage

%%%%%%%%%%%%%%%%%%%%%%%%Fig-2%%%%%%%%%%%%%%%%%%%%%%%%%%%%
%%%%%%%%%%%%%%%%%%%%%%%%%%%%%%%%%%%%%%%%%%%%%%%%%%

\newpage

%%%%%%%%%%%%%%%%%%%%Fig-3%%%%%%%%%%%%%%%%%%%%%%%%%%%%%
%%%%%%%%%%%%%%%%%%%%%%%%%%%%%%%%%%%%%%%%%%%%%%%%%%%%%%%%%%%%%%

\newpage
%%%%%%%%%%%%%%%%%%%%%%%%%% Fig-4 %%%%%%%%%%%%%%%%%%%%%%%%%%%%%%%%%%%%

\newpage

%%%%%%%%%%%%%%%%%%%%%%%%%%%%Fig-5%%%%%%%%%%%%%%%%%%%%%%%%%%%%%%%%%
%%%%%%%%%%%%%%%%%%%%%%%%%%%%%%%%%%%%%%%%%%%%%%%%%%%%%%%%%%%%%%%%

\newpage

%%%%%%%%%%%%%%%%%%%%Fig-6%%%%%%%%%%%%%%
%%%%%%%%%%%%%%%%%%%%%%%%%%%%%

\newpage

%%%%%%%%%%%%%%%%%%%%%%%%Fig-7%%%%%%%%%%%%%%%%%%%%%%%%%%%

%%%%%%%%%%%%%%%%%%%%%%%%%%%%%%%%%%%%%%%%%%%%%%%%%

\newpage

%%%%%%%%%%%%%%%%%%%%Fig-8%%%%%%%%%%%%%
%%%%%%%%%%%%%%%%%%%%%%%%

\newpage

%%%%%%%%%%%%%%%%%%%%%%%% Fig-9 %%%%%%%%%%%%%%%%%%%%%%%%%%%%%

%\newpage

%%%%%%%%%%%%Fig-10%%%%%%%%%%%%%%%%%%%%%%%%%%%%%%%%
%%%%%%%%%%%%%%%%%%%%%%%%%%%%%%%%%%%%%%%%%%%%%%%%%%%

\newpage

%%%%%%%%%%%%%%%%%%%%%% Fig-11%%%%%%%%%%%%%%%%%%%%%%%
%%%%%%%%%%%%%%%%%%%%%%%%%%%%%%%%%%%%%%%%
%\begin{figure}
%    \centering
%    \includegraphics[width=0.5\linewidth]{chi-max-L.pdf}
%    \caption{The log-log plot of $\chi_{max}$ with system sizes $L$ gives the critical exponent ratio of susceptibility as $%\frac{\gamma}{\nu}$=0.532}
%    \label{fig:chi-max-L}
%\end{figure}
 
%%%%%%%%%%%%%%%%%%%%%%%%%%%%%%%%%%%%%%%%%%%%%%%%%

\end{document}